\documentclass[prb,aps,twocolumn,showpacs,superscriptaddress,floatfix]{revtex4}
\usepackage{natbib}
\usepackage{epsfig}
\usepackage{amsmath}
\usepackage{amssymb}
\usepackage{amsfonts}
\usepackage{graphicx}

\providecommand{\U}[1]{\protect\rule{.1in}{.1in}}

\begin{document}

\title{Evidence of quantum phase slip effect in titanium nanowires}
\author{J. S. Lehtinen}
\affiliation{University of Jyv\"{a}skyl\"{a}, Department of Physics, PB 35, 40014 Jyv\"{a}%
skyl\"{a}, Finland}
\author{T. Sajavaara}
\affiliation{University of Jyv\"{a}skyl\"{a}, Department of Physics, PB 35, 40014 Jyv\"{a}%
skyl\"{a}, Finland}
\author{K. Yu. Arutyunov}
\email{Konstantin.Arutyunov@phys.jyu.fi}
\affiliation{University of Jyv\"{a}skyl\"{a}, Department of Physics, PB 35, 40014 Jyv\"{a}%
skyl\"{a}, Finland}
\affiliation{Nuclear Physics Institute, Moscow State University, 119992 Moscow, Russia}
\author{A.~Vasiliev}
\affiliation{Kurchatov Institute, 1, Akademika Kurchatova pl., 123182 Moscow, Russia}
\date{\today}

\begin{abstract}
Electron transport properties of titanium nanowires were experimentally
studied. Below the effective diameter $\lesssim$ 50 nm all samples
demonstrated a pronounced broadening of the $R(T)$ dependencies, which
cannot be accounted for thermal flcutuations. An extensive microscopic and
elemental analysis indicates the absence of structural or/and geometrical
imperfection capable to broaden the the $R(T)$ transition to such an extent.
We associate the effect with quantum flucutuations of the order parameter.
\end{abstract}

\maketitle


Since the early years of experimental studies in superconductivity it has
been noticed that the supercondcunting transition $R(T)$ has always a finite
width. Very often the broadening can be accounted for sample inhomogeneity.
However, soon it became clear that, at least in low dimensional samples, the
transition width remains finite even with the refined material purity and
improved fabrication. The effect has been attributed to fluctuations
typically more pronounced in objects with reduced dimensionality. The finite
resistance $R(T)\sim\exp\left( -F_{0}/k_{B}T\right) $ at a temperature $T$
below the critical temperature $T_{c}$ of a quasi-one-dimensional
superconducting channel with cross section $\sigma$\ has been explained by
the thermal fluctuations of the order paprameter: the so called thermal
activation of phase slips (TAPS), \cite{Langer-Ambegaokar67},\cite%
{McCumber-Halperin70}. Here the condensation energy $F_{0}\sim
B_{c}^{2}\xi\sigma$ of the smallest statistically independent volume $%
\xi\sigma$, where $\xi$\ is the supercondcuting coherence length and $B_{c}$%
\ is the critical magnetic field, competes with the thermal energy $k_{B}T$.
The effect manifests itself only sufficiently close to the critical
temperature, and in extreemly homogeneous samples with micrometer-size
diameter (e.g. pure whiskers) leads to the experimentally observable width
of the $R(T)$ transition of about few mK \cite%
{Lukens-PRL1970-tin-whisker-R(T)}, \cite{Newbower-PRB1972-tin-whisker-R(T)}, 
\cite{Arutyunov-PRB2001-Whisker}. In less homogeneous objects (e.g.
lithographically fabricated nanowires) separation of the impact of the
thermal fluctuations from the trivial inhomogeneity-determined $R(T)$
broadening is rather problematic \cite{Zgirski-PRB2007-Limits}. Nevertheless
with development of nanotechnology \cite{Arutyunov-Patents2007-1Dtechnology}
it became clear that in extreemly narrow superconducting wires, with
diameters $\sim$10 nm, the shape of the $R(T)$ transition by no means can be
explained by sample inhomogeneity or/and thermal fluctuations \cite%
{Arutyunov-PhysRep2008}. The effect has been attributed to quantum
fluctuations, also called - \textit{quantum phase slips} (QPS) - and has
been observed in a rather limited number of experiments studying the
transport properties of ultra-narrow nanowires made of various
superconducting materials: amorphous $MoGe$ \cite{Grabeal-PRL87-QPS-MoGe}, 
\cite{Bezryadin-Tinkham-Nature2000-QPS}, \cite{Lau-Tinkham-PRL2001-QPS}; $In$
and $In-Pb$ \cite{Giordaano-PRL88-QPS}, \cite{Giordaano-PRL89-QPS}, \cite%
{Giordaano-PhysicaB94-QPS}; $Al$ \cite{Altomare-PRL05-QPS}, \cite%
{Zgirski-NanoLett05-QPS}, \cite{Zgirski-PRB08-QPS}. Being not explicitly
attributed to quantum fluctuations, the anomalous $R(T)$ broadening has been
reported in $Pb$ \cite{Sharifi-Dynes-PRL93-1Dwire} and amorphous $InO_{x}$ 
\cite{Johansson-PRL2005-nanowire}. Though the subject of a reliable
experimental confirmation of the quantum fluctuation phenomenon in
quasi-one-dimensional superconductors is still under debates, there is a
consensus in the scientific community, that if exists, it should be observed
in extreemly narrow samples with characteristic diameters $\sim$10 nm.
Unfortunately, at these scales an independent and reliable analysis of a
nanostructure homogeneity is rather problematic. The uncertainty leaves a
room for a critically-oriented scientist to attribute the experimentally
observed deviations of the $R(T)$ superconducting transition shape from the
well-established TAPS model to sample inhomogenities: structural
(impurities, grain boundaries) or geometrical (constrictions). In this paper
we demonstrate that with proper selection of material (superconducting
titanium) the obvious deviations of the $R(T)$ transition shape from the
TAPS model become pronounced already at scales $\lesssim$ 50 nm, which
drammatically simplifies both the sample fabrication and makes the
structural analysis more reliable compared to the 10 nm case.

\begin{figure}[t]
\epsfxsize=7cm\epsffile{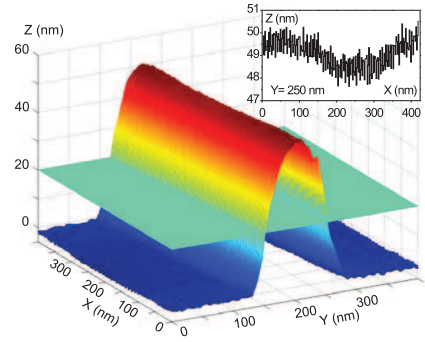}
\caption{(Color online). SPM image of typical part of a titanium nanowire.
Horisonal plane indicates the interface between the metal and the sputtered
Si substrate. The experiemental error in definition of the interface
position provides the main contribution to uncertainty in determination the
sample cross section. Inset: profile of the top part of the sample. }
\label{Fig.1}
\end{figure}

So far a rather limited number of models has been proposed to describe the
impact of quantum fluctuations on transport properties of quasi-one
dimensional superconductors \cite{Arutyunov-PhysRep2008}. Following \cite%
{Zaikin-PRL1997-QPS}, \cite{Zaikin-Uspexi98-QPS}, \cite{Golubev-Zaikin-PRB01}
one may describe the QPS contribution to the effective resistance of a
superconducting wire with length $L$ and cross section $\sigma$ as:%
\begin{equation}
R(T)\simeq b\frac{\Delta(T)S_{QPS}^{2}L}{\xi(T)}\exp(-2S_{QPS})   \label{Eq1}
\end{equation}

where $b$\ is an unimportant constant which remains the same for all
samples; $\Delta(T)$ and $\xi(T)$ are the temperature-dependent
superconducting energy gap and coherence length, respectively. The QPS
action $S_{QPS}=A\left( R_{Q}/R_{N}\right) (L/\xi)$, where $R_{Q}=\hbar/2e=$
6.47 k$\Omega$ is the 'supercondcuting' quantum resistance and $R_{N}$ is
the wire resistance in the normal state. Constant $A$ is of the order of
unit and, unfortunately, cannot be determined more precise within the model 
\cite{Zaikin-PRL1997-QPS}, \cite{Zaikin-Uspexi98-QPS}, \cite%
{Golubev-Zaikin-PRB01}. The mean free path $\ell$ and the constant $A$ are
the two fitting parameter of the QPS model. Note that the mean free path is
not a truly free parameter and with acceptable accuracy can be estimated
from the normal state resistivity as the product $\ell\rho_{N}$ is the
material\ constant. Two other parameters of the model - the critical
temperature\ $T_{c}$ and the normal state resistance $R_{N}$ - are deduced
from the the experimental $R(T)$ dependencies. Recovering that for a 'dirty
limit' superconductor $\ell\ll\xi$ the coherence length $\xi \simeq\sqrt{%
\ell\xi_{0}}$, where the 'clean' coherence length $\xi_{0}\simeq\hbar
v_{F}/\Delta$ and $v_{F}$\ is the Fermi velocity, one may conclude that with
the exponential accuracy $R(T)\sim\exp(-aT_{c}^{1/2}\sigma/\rho _{N})$. As
the Fermi velocity does not vary much between conventional supercondcutors
being of the same order as $v_{F}\simeq1.79\times10^{6}$ m/s for titanium,
the constant $a$\ should be basically material-independent. So far in the
rather limited number of experiments claiming the observation of QPS
phenomena \cite{Grabeal-PRL87-QPS-MoGe}-\cite{Zgirski-PRB08-QPS} the efforts
were mainly concentrated on reducing the sample cross section $\sigma $.
However, the material issue was largely ignored: for the smallest obtainable
dimension $\sigma$, obviously limited by available fabication capabilities,
the ultra-low temperature supercondcutors with high normal state resistivity
are of advantage.

In this paper titanium has been selected as the suitable material for
demonstration of the QPS phenomenon. The crtitical temperature $T_{c}$ of
the titanium nanowires is below 400 mK. The low-temperature resistivity $%
\rho_{N}$ is significantly larger than for the majority of single-element
supercondcutors and varies from $\simeq1.0\times10^{-6}$ $\Omega\times$m for
the 2D films to $\simeq3.2\times10^{-6}$ $\Omega\times$m for the sub-30 nm
nanowires. Utilizing for titanium the product $\ell\rho_{N}\simeq
10\times10^{-16}$ $\Omega\times$m$^{2}$, which slightly varies from
different literature sources \cite{Sanborn-PhysRevB-40-6037-material-data},
one gets that in our nanowires the mean free path $\ell$ is of the order of
1 nm, which correlates with the independent transmission electron microscope
(TEM) analysis indicating the corresponding size of the defect-free areas.
From the technological point of view titanium is an easy-to-deal material.
The nanowires with different lengths $L$ between 1 $\mu$m\ and 100 $\mu$m
were fabricated using conventional lift-off technique: soft-mask e-beam
lithography followed by e-gun evaporation at a residual pressure $\sim10^{-9}
$mBar on naturally oxidized $Si/SiO_{x}$ substrate. The mentioned difference
in normal state resitivity $\rho_{N}$\ between the wide films and the thin
nanowires presumably originates from the pressure gradient between the
bottom of the narrow groove in the resist mask and the rest of the vacuum
chamber. In the former case, being an effective getter material, Ti
'absorbs' the residual gas and the not-completely-evacuated organics leading
to formation of a dirtier sample. After the analysis with the scanning
electron and scanning probe microscopes (SEM and SPM, respectively), the
samples showing no obvious defects were cooled down in $^{3}$He$^{4}$He
dilution refrigerator down to temperatures $T\simeq$ 50 mK. Conventional
four-probe DC and low-frequency ($<$ 20 Hz) lock-in AC techniques were used
to measure the $R(T)$ dependencies. Special care was taken not to overheat
the samples. Very low excitation currents down to 30 pA were used to ensure
the linear response. The refrigerator and the front-end battery powered
analogue pre-amplifiers were located inside the electromagnetically shielded
room being connected to the external electronics via low-pass RC filters. An
extensive multi-stage RLC filtering of the signal lines inside the
refrigerator effectively reduced the impact of the noisy electromagnetic
environment. In a separate experiment using the same measuring set-up, the
increase of the effective electron temperature $T_{e}$ deduced from the
shape of the $I-V$ dependence of a normal-insulator-supercondcutor junction 
\cite{Pekola-Arutyunov-PhysicaB2000-NIS}, was found to be at the level $%
\delta T_{e}\lesssim$ 20 mK above the base temperature $T$ \cite%
{PhysRevB-83-104509-Arutyunov-NIS}.

\begin{figure}[t]
\epsfxsize=7cm\epsffile{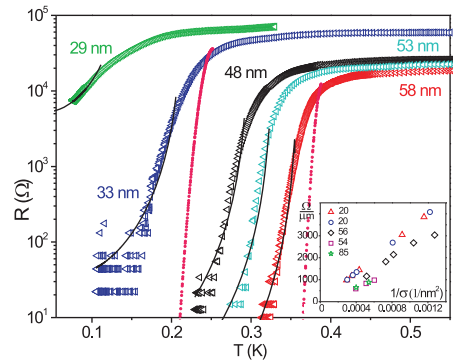}
\caption{(Color online) Resistance vs. temperature for the same titanium
nanowires with length L= 20 $\protect\mu$m and progressively reduced
effective diameter $\protect\sqrt{\protect\sigma}$ indicated in the plot and
specified with accuracy $\pm$2 nm. Inset: normal state resistance per
unitlength vs. inverse cross sections for several sample swith different
length L indicated in $\protect\mu$m. }
\label{Fig.2}
\end{figure}

Both the co-deposited 2D films with characteristic thickness $d\simeq$40 nm
and the as-fabricated relatively 'wide' $w>$ 60 nm nanowires showed an
abrupt superconducting transition at a critical temperature $T_{c}\simeq$
400 mK (Figs. 2 and 3). The experimentally measurable width of the $R(T)$
transition $\delta T_{c}\simeq$ 20 mK of these 'thick' samples can be
qualitatively understood in terms of the TAPS model and the inevitable mild
inhomogeneity, e.g. few percent variation of the cross section \cite%
{Zgirski-PRB2007-Limits}. After each session of the $R(T)$ measurments the
samples were subjected to the low-energy ion etching to reduce the wire
effective diameter $\sqrt{\sigma }\equiv\sqrt{dw}$ with steps as small as $%
\simeq$1 nm \cite{Savolainen-APA2004-IBE}, \cite%
{Zgirski-Nanotechnology2008-IBE}. Due to the re-deposition of the sputtered
material the inevitable imperfections are smoothed-out resulting in the
surface roughness of the processed samples approaching $\pm$ 1 nm (Fig. 1).
The method enables the study of a size-dependent phenomenon on a same
structure with progressively reduced characteristic dimension elimintating
the artefacts of the samples fabricated in different experimental runs.
Further decrease of the nanowires cross sections leads to (i) reduction of
the critical temperature $T_{c}$, and (ii) broadening of the $R(T)$
transition (Figs. 2 and 3). The first effect is well-known for low
dimensional superconductors \cite{Arutyunov-PhysRep2008}, though the origin
of the phenomenon is still under debates \cite{Shanenko-PRB2006-Tc}. As the
size dependence of $T_{c}$ is observed not only in nanowires, but also in
wide 2D films, we believe that the effect is not related to the essentially
1D phenomenon under discussion - phase slips (thermal or quantum).

\begin{figure}[t]
\epsfxsize=7cm\epsffile{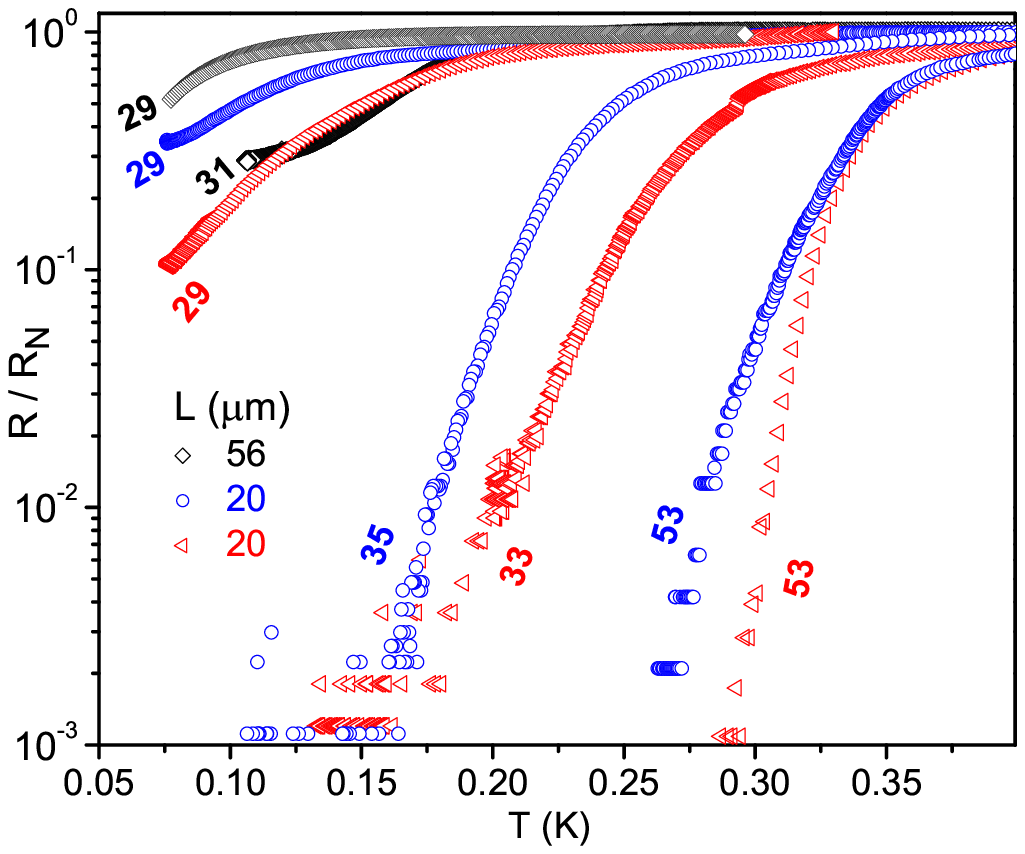}
\caption{(Color online). Normalized resistance vs. temperature for three
different nanowires of length L. The effective diameters $\protect\sqrt{%
\protect\sigma}$\ are indicated in the plot in nm and are specified with
accuracy $\pm$2 nm. }
\label{Fig3}
\end{figure}

Homogeneity of samples is the key point in interpretation of experimental $%
R(T)$ data within any model derived for a homogeneous superconducting
channel of uniform cross section $\sigma$. Critically oriented reader might
always argue that the observed broad $R(T)$ transition is the result of a
trivial sample inhomogeneity: either structural (e.g. local variation of the
critical temperature along the wire) or geometrical (e.g. constrictions).
The two cases should be analysed separately. Let us first consider the
structural inhomogeneity. Obviously all real samples do have a certain level
of structural inhomogeneity originating from various sources: grain
boundaries, finite size of the sample, proximity effect at the interface
with wider parts of the structure, etc. We would like to note that all our
samples, including the thinnest wires with a typical sheet resistance $%
R_{\square}\lesssim$ 200 $\Omega$, are still comfortably on the metal side
of a 'dirty' titanium. Formation of a network of weakly coupled metal grains
and the corresponding Coulomb effects have been observed in deliberately
oxidized titanium films with the sheet resistance $R_{\square}$ exceeding
few k$\Omega$ \cite{Schollmann-APL2000}, \cite%
{Johannson-Haviland-PhysicaB-2000}. All our samples above the critical
temperature demonstrate $I-V$ dependencies without any non-linearities,
which otherwise might indicate the existence of weak links (Fig. 4), and the
dependence of the normal state resistance $R_{N}$ on diameter follows the
expected Ohm's law (Fig. 2, inset). If the transport properties of the
nanowires would be determined by the presence of the weak links, then at low
temperatures the non-linear $I-V$ characteristics should be
sample-dependent. For example, various parts of a long multi-terminal
nanowire should demonstrate some 'fingerprint' features related to formation
of weak links particular for each sub-sample, which is not the case (Fig.
4). 
\begin{figure}[t]
\epsfxsize=7cm\epsffile{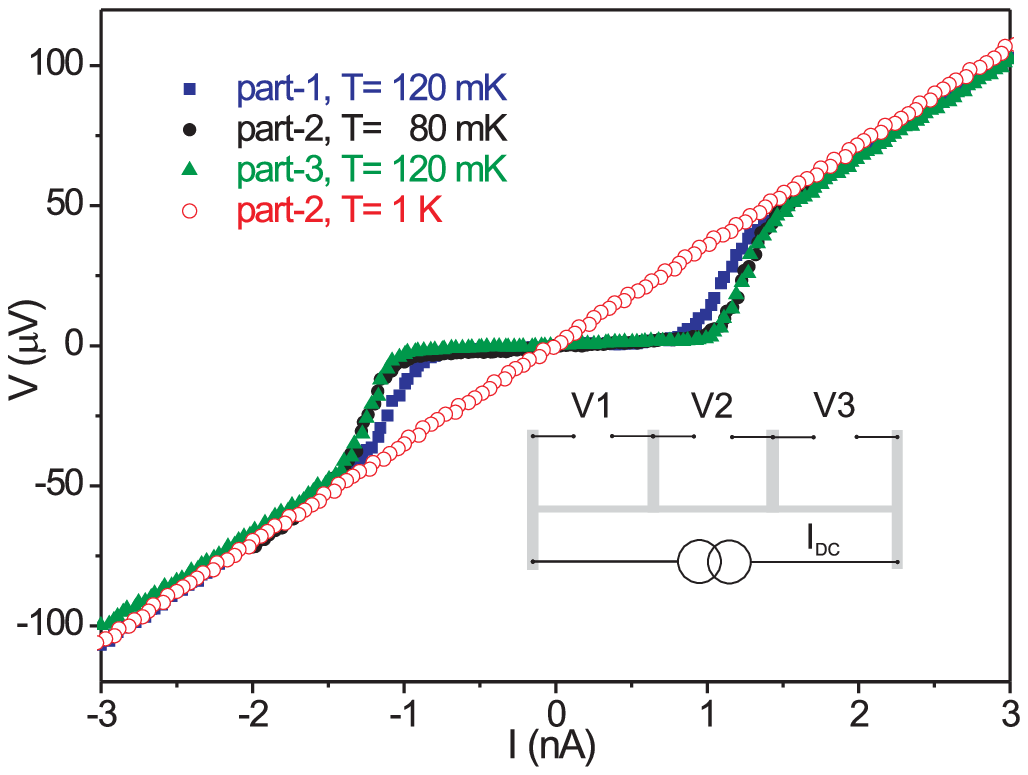}
\caption{(Color online). $V(I)$ characteristics of the three neighbouring
parts with equal lengths $L$ = 20 $\protect\mu$m of the same nanowire with
the effective diameter $\protect\sqrt{\protect\sigma}$ = 38 $\pm$ 2 nm .
Above the critical temperature $T_{c}\simeq$ 300 mK within the scale of the
image the Ohmic $V(I)$ characteristics ($\circ$) are quantitatively
indistinguishable between the different parts. Below the critical
temperature the $V(I)$ dependencies are also very similar ($\blacksquare$,$%
\bullet$,$\blacktriangle$). Note the absence of the true zero resistance
state below the 'critical current' value $\simeq$ 1.2 nA. Inset shows the
layout of the sample and the measurement.}
\label{Fig4}
\end{figure}

An extensive TEM analysis (Fig. 5) cannot reveal any suspicious structural
imperfections inside the metal matrix: the material bulk looks exactly the
same for the sputtered and for the non-sputtered samples. The
polycrystalline nanostructures consist of compactly packed grains with the
average size of defect-free area $\simeq$ 3 nm, which correlate well with
the the best-fit value $\ell\simeq$ 1 nm used in calculations. The elemental
depth profiles (Fig. 6) were determined by means of Time-of-Flight Elastic
Recoil Detection Analysis (TOF-ERDA) \cite{Putkonen-ABC2005} using 8.015 MeV 
$^{35}$Cl$^{4+}$ incident ions. The areal density of oxygen at the surface
(1.7$\times$10$^{16}$ atoms/cm$^{2}$) and at the interface (5.5$\times$10$%
^{15}$ atoms/cm$^{2}$) corresponds to the thickness of about 1.9 nm and 0.8
nm, respectively, for the surface TiO$_{2}$ and the boundary with SiO$_{2}$%
/Si with the corrersponding density 4.0 g/cm$^{3}$ and 2.2 g/cm$^{3}$. The
thickness of the material with the high concentration of oxygen correlates
well with the high-resolution TEM anaysis data (Fig. 5). The bulk
concentration of oxygen $\simeq$ 0.4 at. \% inside the titanium matrix was
determined by comparison of the experimental energy spectra with the ones
obtained by Monte Carlo simulations \cite{Arstila-NIMB2001}. Given that the
average microcrystall size is $\sim$ 3 nm, one can easily estimate that
inside the titanium bulk there is less that one oxygen atom per defect
boundary. Concentration of other than oxygen elements inside the titanium
matrix was found to be even smaller. The observation eliminates the
possibility of the weak link(s) formation due to non-metallic grain
interfaces capable to block the metal-to-metal supercurrent. To summarize,
all available at our disposal methods of analysis - SEM, TEM, SPM and
TOF-ERDA - give us a confidence to state that (i) our nanostructures are as
homogeneous, as a conventional thin film titanium can be; and (ii) - what is
even more important - our method of reduction of the nanowire cross section
by the low energy ion sputtering does not introduce new defects. At
acceleration energies $\simeq$ 1 keV the penetration depth of the $Ar^{+}$
ions inside the titanium matrix is below 3 nm making the method virtually
non-destructive: the thickness of the ion-damaged layer is comparable to the
thickness of the naturally grown oxide. Hence, if there are some inevitable
'intrinsic' structural defects, they cannot appear in thinner samples
contributing to broadening of the $R(T)$ transition. If to assume that there
exist some mysterious and undetectable mechanism which degrades the wire
homogeneity with reduction of its diameter (e.g. recovery of hidden in the
bulk cavern(s) or/and grain boundaries, blocking the supercurrent), it is
reasonable to assume that the mechanism should manifest itself individually
for each particular sample leading to a unique 'fingerprint' on the $R(T)$
and the $I-V$ dependencies. However, a large statistics of samples states
the opposite: the shape of the broadened $R(T)$ transitions is universal and
reproducible for different samples of the same effective diameter (Fig. 3).

\begin{figure}[t]
\epsfxsize=7cm\epsffile{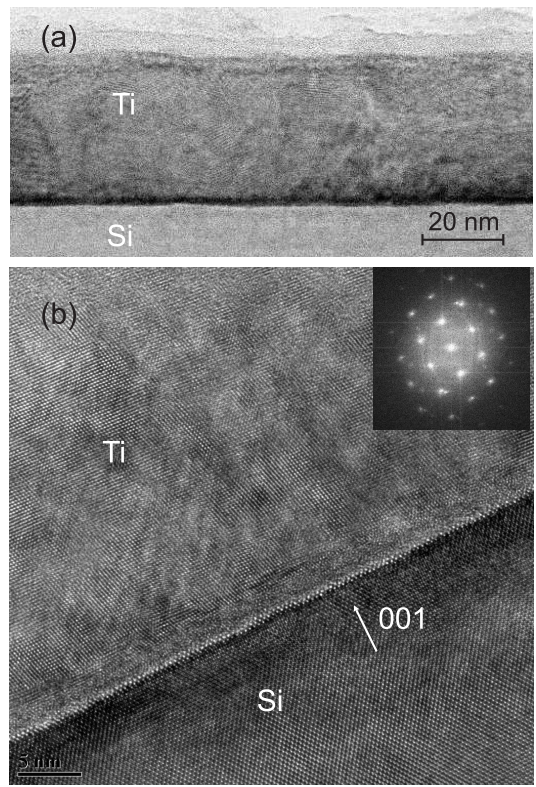}
\caption{(a) Bright field low-resolution TEM image of the cross section of a
typical 35 nm thick Ti film on Si substrate. (b) Bright field
high-resolution TEM image of the Ti / Si interface. Inset: Fast Fourrier
Transform (FFT) indicates the single crystall nature of the metal grains
forming the film. }
\label{5}
\end{figure}

Now let us turn to the alternative critics dealing with geometrical
inhomogeneity: the inevitable variation of a nanowire cross section. One
might argue that the $R(T)$ broadening originates from the well-known TAPS
mechanism: due to the strong exponential dependence $R(T)\sim\exp(-\sigma)$
the contributions coming from the parts with different cross sections $%
\sigma_{local}$ wash-out the otherwise sharp $R(T)$ dependence. First, we
would like to note that with reduction of the wire dimensions (by ion
sputtering) the shape of the $R(T)$ transition becomes more consistent
between the samples with close values of the cross sections (Fig. 3). The
observation supports the earlier statement that the ion beam etching
polishes the surface making the sample geometrically more uniform: the
as-fabricated 'wide' nanowires have larger variation of the cross section $%
\sigma$ compared to the thinner ('polished') samples. Here we would like to
stress, that the uncertainty of the effective diameter specified in Figs. 2
and 3 reflects the experimental error in determination of the cross section
and not the actual roughness of the surface, which for the multiply
sputtered samples does not exceed $\pm$1 nm (inset in Fig.1). The error
mainly comes from the uncertainty in determination of the position of the
interface between the metal and the sputtered substrate (Fig. 1). In SEM the
contrast between the two light materilas - Ti and Si - is not sufficent to
determine the postion of the interface with accuracy better than $\pm$2 nm.
While the tip deconvolution effect, typical for SPM analysis of essentailly
3D nanostructures, leads to the basically same uncertainty. Summarizing, we
would like to state that by standards of the modern nanotechnology and
microscopic analysis our nanowires are so 'large' that it is almost
impossible to overlook a pronounced constiction. Even in the worst case
scenario, in the sputtered samples the variation the cross sections along
the wire is below $\pm$10\%. The contribution of such a moderate geometrical
imperfection on the shape of the $R(T)$ transition determined by thermal
fluctuations has been analysed \cite{Zgirski-PRB2007-Limits}: by no means it
can account for the experimentally observed pronounced broadening of the $%
R(T)$ dependencies. One may come to the same conclusion just analysing Fig.
2: the reduction of the wire average diameter from 58 nm to 33 nm leads to a
negligible broadeneing of the TAPS fits (dashed lines) compared to the
experimental data (symbols). For sufficiently narrow nanowires no realistic
set of fitting parameters of the TAPS model can explain the broad
experimental $R(T)$ dependencies (e.g. Fig. 2, symbols).

\begin{figure}[t]
\epsfxsize=7cm\epsffile{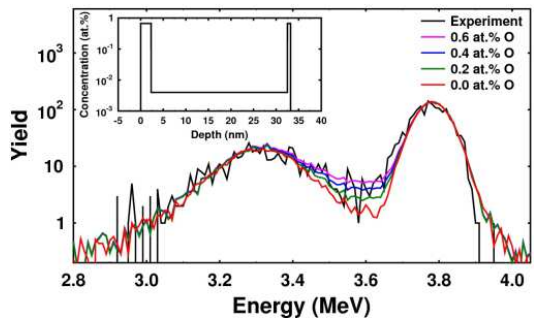}
\caption{(Color online). Experimental and Monte Carlo simulated energy
spectra for oxygen recoils from the 33 nm thick Ti film and Si substrate. In
the simulation the surface and interface oxygen contents were fixed, while
the concentration in the film was varied. The best fit can be obtained with
0.4 at.\% concentration. The full depth profile of oxygen is shown in the
insert.\protect\bigskip}
\label{6}
\end{figure}

On the contrary, the QPS model provides reasonable agreement with the
experiment (Fig. 2, solid lines). In simulations of the theoretical
QPS-governed $R(T)$ dependencies (Eq. \ref{Eq1}) for each sample (cross
section) the best fit critical temperature $T_{c}$ corresponds to the onset
of superconductivity, the mean free path $\ell\simeq$ 1 nm and the numerical
constant $A\simeq$ 0.3 were kept as free parameters. At temperatures $T\ll
T_{c}$ the negative magnetoresistance of about few percent has been observed
in the thinnest samples at very small magentic fields $\lesssim$ 3 mT.
Similar effect has been earlier reported in ultra-narrow lead \cite%
{Xiong-Dynes-PRL97} and aluminum \cite{Zgirski-PRB08-QPS} nanowires. The
origin of the pheonomenon is not yet clear. One alternative \cite%
{Arutyunov-PhysicaC2008-nMR} employs possible formation of a charge
imbalance region accompanying each phase slip event. This non-equilibrium
region, if exists, would provide dissipation outside the core of a phase
slip. Within a certain range of (small) magnetic fields the corresponding
Ohmic contribution can be suppressed by the magnetic field more effective
than the superconducting gap, resulting in the negative magnetoresistance.
However, so far the validity of the charge imbalance concept has been only
demonstrated at temperatures sufficiently close to $T_{c}$ and its
applicability to QPS is by no means obvious. The charge imbalance scenario
responsible for the negative magnetoresistance \cite%
{Arutyunov-PhysicaC2008-nMR} still requires a solid theoretical
justification.

It should be noted that the both models (QPS and TAPS) were derived assuming
that the phase slips are the 'rare' events. In other words, the justified
comparison with experiment is valid only in the limit $R(T)\ll R_{N}$. The
QPS effect is an essentially low temperature phenomenon when quantum
fluctuations of the order parameter dominate over the thermal fluctuations,
with the latter being important only sufficiently close to the critical
temperature $T\longrightarrow T_{c}$. Contrary to the QPS, extrapolation of
the TAPS mechanism down to lower temperatures violates the Ginzburg
criterion $\left( T_{c}-T\right) /T_{c}\ll$1 of the model applicability \cite%
{Meidan-PRL07-R(T)}. This is an additional ('theoretical') argument why the
broad experimental $R(T)$ transitions (Figs. 2 and 3) cannot be explained in
terms of thermal fluctuations \cite{Langer-Ambegaokar67}, \cite%
{McCumber-Halperin70}, even if one would assume in those samples the
presence of extended and unrealistically narrow constrictions $\sqrt{\sigma }%
\sim$1 nm (overlooked in all microscopes!).

In conclusion, we have studied titanium nanowires with progressively reduced
cross sections. An extensive microscopy and elemental analysis revealed no
obvious structural or geometrical imperfections. Neither the normal state,
nor the supercondcuting transport properties provided any signature of a
non-Ohmic behavior to be associated with 'hidden' structural defects. The
thickest samples demonstrated relatively sharp $R(T)$ supercondcuting
transitions with the shape which can be qualitatively understood by the
model of thermally activated phase slips (TAPS) \cite{Langer-Ambegaokar67}, 
\cite{McCumber-Halperin70} and the inevitable mild inhomogeneity of the
samples \cite{Zgirski-PRB2007-Limits}. However for the nanowires with
diameters $\lesssim$ 50 nm the width of the $R(T)$ transition broadens well
above the limits which can be explained by the TAPS model with a realistic
set of parameters. For the thinnest samples with diameters $\lesssim$ 30 nm
the temperature dependence of the experimentally measured resistance $R(T)$
is very weak and does not extrapolate to zero at $T\longrightarrow0$. By the
standards of modern nanotechnology and microscopic analysis the structures
are so 'large' that speculations about trivial inhomogeneity overlooked in
SET, TEM and SPM microscopes could be ruled out with a high level of
confidence. We associate the pronounced broadening of the $R(T)$ transitions
with quantum fluctuations of the order parameter - the so-called quantum
phase slips (QPS) \cite{Zaikin-PRL1997-QPS}, \cite{Zaikin-Uspexi98-QPS}, 
\cite{Golubev-Zaikin-PRB01}. Additionally to the importance for the basic
knowledge about nanoscale supercondcutivity, the subject of quantum
fluctuations is expected to lead to a new class of devices: quantum standard
of electric current \cite{Mooij-Nazarov-NatPhys06}, qubit \cite%
{Mooij-NJP2005-QPS-qbit} and various QPS-based systems \cite%
{Nazarov-PRL2011-QPSoscillator}, \cite{Nazarov-PRB2011-QPSjunction}.

\bigskip

\begin{acknowledgments}
The authors would like to acknowledge support of TEKES project DEMAPP, Grant
2010-1.5-508-005-037 of Russian Minsitry of Education and Research and thank
T. T. Hongisto for valuable discussions and L. Leino for help with SPM
analysis.
\end{acknowledgments}

\end{document}